\documentstyle[amstex,amssymb]{article}

\begin{document}

\title{The ''Cameo Principle'' and the Origin of Scale-Free Graphs in Social
Networks}
\author{Ph.~Blanchard$^{1}$ and T.~Kr\"uger$^{2}$ \\
Universit\"{a}t Bielefeld\\
Faculty of Physics$^{1}$\\
and Faculty of Mathematics$^{2}$\\
D-33615 Bielefeld, Germany\\
---------------------\\
\\
{\em Dedicated to the memory of Per Bak}\\
\\
----------------------}
\date{\today }
\maketitle

\begin{abstract}
We formulate a simple edge generation rule based on an inverse like mass
action principle for random graphs over a structured vertex set. We show
that under very weak assumptions on this structure one obtains a scale free
distribution for the degree. We furthermore introduce and study a ''my
friends are your friends'' local search principle which makes the clustering
coefficient large.
\end{abstract}

\section{ Introduction}

Random graphs with a scale free distribution for the degree seem to appear
very frequently in a great variety of real life situations like the
WorldWideWeb, the Internet, social networks, linguistic networks, citation
networks and biochemical networks. An excellent recent survey is the Article
by Albert and Barabasi \cite{1}.

Despite the large number of articles dealing with specific properties of
such networks little is known about origin and formation principles of such
graphs in the general framework. Most popular became the model of Albert
and\ Barabasi \cite{3} where graphs grow by successively adding a new vertex
say $x$ at each time step and forming edges between this new vertex and the
already existing ones by assuming that the probability that $x$ forms an
edge with a vertex $y$ whose degree is $k$ is about $const\cdot \frac{k}{N}$
where $N$ is the total number of vertices at that time. This gives a scale
free graph although a very homogeneous one with little room for
incorporating additional structures. The principle used in this
construction, and the many variants of it, could be seen as a
degree-mass-action principle since the degree acts here as a kind of
positive affinity parameter for alignment of new vertices. Of course
knowledge about the degree of a vertex is in real networks seldom available
for other vertices. For instance in the social networks of \ friendships an
individual certainly does not make the formation of a new friendship
dependent of checking how many friends the potential new candidate already
has (not to mention that it would be rather difficult to estimate this
number). So we seek for an edge formation principle more related to a
pre-given structure on the vertex set. We assume that this structure is
specified by a real positive parameter $\omega \in {\Bbb R}^{+}$- like
social importance, money or beauty- with a given probability distribution $%
\varphi \left( \omega \right)$. Furthermore we think of an edge between
vertices $x$ and $y$ as the result of a directed choice made by either $x$
or $y$ (symbolized by $x\rightarrow y$ or $y\rightarrow x)$. This seems to
be very reasonable since in many real life networks edges are formed that
way. Note that although the edge creation is a directed process we consider
the resulting graph as an undirected one since for the majority of relevant
transmission processes on the network the original orientation of an edge is
irrelevant. The crucial assumption about the pairing probability is now a
kind of inverse mass-action-principle. Namely that the probability that a
vertex $x$ which makes a choice chooses a given vertex $y$ with parameter $%
\omega \left( y\right) $ is proportional to $\left[ \varphi \left( \omega
\left( y\right) \right) \right] ^{-\alpha }\cdot \frac{1}{N}$ with some $%
\alpha >0$ and $N$ being the total number of vertices. Due to this principle
it is not the actual value $\omega $ of a vertex $y$ which is relevant for
the pairing probability but rather it's relative frequency of appearance in
the ''population''. This is in a certain sense a market rule -- the more
rare a property is, the more attractive it becomes for others. Of course the
property associated with $\omega $ has to be positive in perception. We
think that this ''Ansatz''\ captures the essence of many real life network
formation processes.

Instead of using a technical name for the principle like ''inverse mass
action'' we want to call it the ''Cameo-principle'' having in mind the
attractiveness, rareness and beauty of the small medallion with a profiled
head in relief called Cameo. And it is exactly their rareness and beauty
which gives them their high value.

Our basic assumptions can be summarized as follows :

\begin{itemize}
\item  The parameter $\omega $ is independent identically distributed
(i.i.d.) over the vertex set with a smooth monotone decreasing density
function $\varphi \left( \omega \right) $

\item  Edges are formed by a sequence of {\em choices}. By a {\em choice} we
mean that a vertex $x$ chooses another vertex, say $y$ , to form an edge
between $y$ and $x.$ A vertex can make several choices. All choices are
assumed to be independent of each other.

\item  If $x$ makes a choice the probability of choosing $y$ depends only on
the relative density of $\omega \left( y\right) $ and is of the form 
\begin{equation}
\Pr \left\{ x\rightarrow y\mid \omega \left( y\right) \right\} \sim \left[
\varphi \left( \omega \left( y\right) \right) \right] ^{-\alpha }\cdot \frac{%
1}{N};\alpha \in \left( 0;1\right) 
\end{equation}

\item  A pre-given outdegree distribution determines the number of choices
made by the vertices. The total number of choices (and therefore the number
of edges) is assumed to be about $const\cdot $ $N.$
\end{itemize}

\medskip

The striking observation under the above assumptions is now the emergence of
a scale-free degree distribution independent of the choice of the $\omega-$
distribution. Furthermore it can be shown that the exponent in the degree
distribution becomes independent of $\varphi \left(\omega \right) $ if the
tail of $\varphi $ decays faster then any power law. The precise formulation
of the above statements is given in theorem 1 and 2 in the next section.

\section{Analysis of the model}

Let $V_{N}=\left\{ 1,...,N\right\} $ be the vertex set of a random graph
space. We are mainly interested in the asymptotic properties for $N$ being
very large. We assign i.i.d. to each element $x$ from the set $V_{N}$ a
continuous positive real random variable (r.v.) $\omega \left( x\right) $
taken from a distribution with density function $\varphi \left( \omega
\right) $. The variable $\omega $ can be interpreted as a parametrization\
of $V_{N}$. For a set $C_{\omega _{0},\omega _{1}}=\left\{ x:\omega \left(
x\right) \in \left[ \omega _{0},\omega _{1}\right] \right\} $ we obtain $%
{\Bbb E}\left( \sharp C_{\omega _{0},\omega _{1}}\right) =N\cdot
\int\limits_{\omega _{0}}^{\omega _{1}}\varphi \left( \omega \right) d\omega 
$. Without loss of generality we always assume $\varphi >0$ on $\left[
0,\infty \right)$. As a technical assumption we will furthermore need the
following:

\smallskip
{\bf Assumption 1}: $\varphi \in C^{2}\left( \left[ 0,\infty \right) \right) 
${\em \ and the second derivatives }$D^{2}\left( \varphi ^{\mu }\right) $%
{\em \ have no zeros for }$\left| \mu \right| \in \left( 0,\mu _{0}\right) $%
{\em \ and }$\omega >\omega _{0}\left( \mu \right) ${\em \ \ (this is just a
monotonicity assumption on the tail of }$\varphi ${\em )}.

Edges are created by a directed process in which the basic events are
choices made by the vertices. All choices are assumed to be i.i.d. The
number of times a vertex $x$ makes a choice is itself a random variable
which may depend on $x$. We call this r.v. $d_{out}\left( x\right) $. The
number of times a vertex $x$ was chosen in the edge formation process is
called the indegree $d_{in}\left( x\right) .$ Each choice generates a
directed edge. We are mostly interested in the corresponding undirected
graph. If we speak in the following about outdegree and indegree we refer
just to the original direction in the edge formation process. Let $p_{\omega
}=\Pr \left\{ x\rightarrow y\mid \omega =\omega \left( y\right) \right\} $
be the basic probability that a vertex $y$ with a fixed value of $\omega $
is chosen by $x$ if $x$ is about to make a choice. For a given realization $%
\xi $ of the r.v. $\omega $ over $V_{N}$ we assume: 
\begin{equation}
p_{\omega }\left( \xi ,N\right) =\frac{A\left( \xi ,N\right) }{\left[
\varphi \left( \omega \right) \right] ^{\alpha }}\cdot \frac{1}{N}
\end{equation}

\noindent
where $\alpha \in \left( 0,1\right) $ and $A\left( \xi ,N\right) $ is a
normalization constant. It is easy to see that the condition $%
\int\limits_{0}^{\infty }\left[ \varphi \left( \omega \right) \right]
^{1-\alpha }d\omega <\infty $ is necessary and sufficient to get $A\left(
\xi ,N\right) \rightarrow A>0$ for $N\rightarrow \infty $ where convergence
is in the sense of probability. Therefore we need $\alpha <1.$ One might
argue that the choice probabilities should depend more explicitly on the
actual realization $\xi $ of the r.v. $\omega $ over $V_{N}$ -not only via
the normalization constant. The reason not to do so is twofold. First it is
mathematically unpleasant to work with the empirical distribution of $\omega 
$ induced by the realization $\xi $ since one had to use a somehow
artificial $N-$ dependent coarse graining. Second the empirical distribution
is not really ''observed''\ by the vertices (having in mind for instance
individuals in a social network). What seems to be relevant is more the
common believe about the distribution of $\omega$. In this sense our
setting is a natural one. To keep the analysis simple we will first assume
in section 2.1. that the r.v. $d_{out}\left( x\right) $ is constant and
equal to $k_{0}$ for all $x.$ We will discuss the situation of a variable $%
d_{out}$ in section 2.2. Since we want to show that under the above
conditions on $\varphi $ we obtain a power law distribution for the degree
we have to say a few words about the notion of power-law distribution in our
framework. We say that a discrete probability distribution $\psi $ on ${\Bbb %
N}$ is of power law type with exponent $\beta $ if 
\begin{equation}
\psi \left( k\right) =\frac{1}{k^{\beta +o\left( 1\right) }}\text{ for }k>1\ .
\label{1000}
\end{equation}
This means that our concept of power law distribution is an asymptotic
one. We think that this is a very reasonable notion since in any empirical
observation one can see only an approximation to a power law.

Before starting with the proof of our claims we would like to mention that
the emergence of a power law distribution in our setting is not so
surprising as it might seem at the first look. The situation is best
explained in form of an example. Let $\varphi \left( \omega \right) =C\cdot
e^{-\omega }$ and define a new variable $\omega ^{\ast }=\frac{1}{\left[
\varphi \left( \omega \right) \right] ^{\alpha }}=\frac{e^{\omega \alpha }}{%
C^{\alpha }}$ . The new variable $\omega ^{\ast }$ can be seen as the
effective parameter to which the vertex choice process applies. What is the
induced distribution of $\omega ^{\ast }$ ? With $F\left( z\right) =\Pr
\left\{ \omega ^{\ast }<z\right\} $ we obtain 
\begin{equation}
F\left( z\right) =\int\limits_{0}^{\frac{1}{\alpha }\ln C^{\alpha }\cdot
z}\varphi \left( \omega \right) d\omega =-z^{\frac{-1}{\alpha }}-C
\end{equation}
and therefore the $\omega ^{\ast }-$ distribution $\phi \left( \omega ^{\ast
}\right) =\frac{1}{\alpha }\cdot \frac{1}{\left( \omega ^{\ast }\right) ^{1+%
\frac{1}{\alpha }}}$. This is a power law distribution with an exponent
depending only on $\alpha .$

\subsection{The case of constant outdegree ( $d_{out}\left( x\right) =k_{0}$
)}

The main result of the paper is contained in the following

\medskip
{\bf Theorem: }

{\bf i) }Let $\varphi \left( \omega \right) =\frac{1}{\omega^{\beta +o\left(
1\right) }}$ with $\beta >2.$

\ \ \ Then $\lim\limits_{N\rightarrow \infty }\Pr \left\{ d\left( x\right)
=k\mid x\in V_{N}\right\} =\frac{1}{k^{1+\frac{1}{\alpha }-\frac{1}{\alpha
\beta }+o\left( 1\right) }}.$

{\bf ii) }Let $\varphi \left( \omega \right) $ fulfill the assumption 1.

\ \ \ Then $\lim\limits_{N\rightarrow \infty }\Pr \left\{ d\left( x\right)
=k\mid x\in V_{N}\right\} =\frac{1}{k^{1+\frac{1}{\alpha }+o\left( 1\right) }%
}.$

\medskip
{\bf Remark: }The theorem says, that under assumption 1 the degree
distributions for the finite random graphs of our model converge for $%
N\rightarrow \infty $ to a power law distribution in the sense of formula (%
\ref{1000}).

\medskip
{\bf Proof:} Since by assumption the outdegree is constant and $d\left(
x\right) =d_{in}\left( x\right) +d_{out}\left( x\right) =d_{in}\left(
x\right) +k_{0}$ asymptotically almost surely, we have to consider only the
indegree distribution. First we compute the expectation (in the limit $%
N\rightarrow \infty $): 
\begin{equation}
{\Bbb E}\left( d_{in}\left( x\right) \mid \omega =\omega \left( x\right)
\right) =\frac{k_{0}\cdot A}{\left[ \varphi \left( \omega \right) \right]
^{\alpha }}
\end{equation}
and therefore 
\begin{equation}
{\Bbb E}\left( d_{in}\left( x\right) \right) =k_{0}\cdot A\cdot \int \left[
\varphi \left( \omega \right) \right] ^{1-\alpha }d\omega\ .
\end{equation}
For a $x$ with fixed $\omega =\omega \left( x\right) $ we get for the
indegree distribution $\psi _{\omega }\left( k\right) =\Pr \left\{
d_{in}\left( x\right) =k\mid \omega =\omega \left( x\right) \right\} $ a
binomial distribution (here we count also multiple edges but the probability
for their appearance vanishes for large $N$) 
\begin{equation}
\psi _{\omega }\left( k\right) =\binom{k_{0}N}{k}\left( \frac{A\left( \xi
,N\right) }{\left[ \varphi \left( \omega \right) \right] ^{\alpha }N}\right)
^{k}\left( 1-\frac{A\left( \xi ,N\right) }{\left[ \varphi \left( \omega
\right) \right] ^{\alpha }N}\right) ^{k_{0}N-k}  \label{5}
\end{equation}
which converges in the limit of large $N$ to a Poisson distribution: 
\begin{equation}
\psi _{_{\omega }}\left( k\right) =\frac{c_{\omega }^{k}}{k!}e^{-c_{\omega }}
\end{equation}
with $c_{\omega }=\frac{k_{0}\cdot A}{\left[ \varphi \left( \omega \right) %
\right] ^{\alpha }}$ . Finally we obtain for the unconditional limiting
probability $\psi \left( k\right) =\lim\limits_{N\rightarrow \infty }\Pr
\left( d_{in}\left( x\right) =k\right) $ the expression 
\begin{equation}
\psi \left( k\right) =\int \varphi \left( \omega \right) \frac{c_{\omega
}^{k}}{k!}e^{-c_{\omega }}d\omega\ .
\end{equation}
The main contribution to $\psi \left( k\right) $ comes from a rather small
interval of $\omega $ -values - called $I_{ess}\left( k\right) $ -- with the
property that for $\omega _{0}\in I_{ess}\left( k\right) $ the expectation $%
{\Bbb E}\left( d_{in}\left( x\right) \mid \omega _{0}=\omega \left( x\right)
\right) $ is of order $k$. The rapid decay of the Poisson distribution will
guarantee that the remaining parts of the integral become arbitrary small
for large $k$. To fill in the details let 
\begin{equation}
I_{ess}\left( k,\gamma \right) =\left[ \varphi ^{-1}\left( \left( \frac{%
k_{0}\cdot A}{k-k^{\frac{1}{2}+\gamma }}\right) ^{\frac{1}{\alpha }}\right)
,\varphi ^{-1}\left( \left( \frac{k_{0}\cdot A}{k+k^{\frac{1}{2}+\gamma }}%
\right) ^{\frac{1}{\alpha }}\right) \right]
\end{equation}
with $0<\gamma \ll 1$. To estimate $\psi _{ess}\left( k\right)
=\int\limits_{I_{ess}\left( k,\gamma \right) }\varphi \left( \omega \right) 
\frac{c_{\omega }^{k}}{k!}e^{-c_{\omega }}d\omega $ we define a new variable 
$\varepsilon $ in the interval $\left[ -k^{\frac{1}{2}+\gamma },k^{\frac{1}{2%
}+\gamma }\right] $ by $\omega =\varphi ^{-1}\left( \left( \frac{k_{0}\cdot A%
}{k+\varepsilon }\right) ^{\frac{1}{\alpha }}\right) $ and transform the
integral into 
\begin{equation}
\psi _{ess}\left( k\right) =\int\limits_{-k^{\frac{1}{2}+\gamma }}^{k^{\frac{%
1}{2}+\gamma }}\frac{-\left( k_{0}\cdot A\right) ^{\frac{1}{\alpha }}}{%
\alpha \left( k+\varepsilon \right) ^{1+\frac{1}{\alpha }}}\cdot \frac{%
\varphi \circ \varphi ^{-1}\left( \left( \frac{k_{0}\cdot A}{k+\varepsilon }%
\right) ^{\frac{1}{\alpha }}\right) }{\left( D\varphi \right) \left( \varphi
^{-1}\left( \left( \frac{k_{0}\cdot A}{k+\varepsilon }\right) ^{\frac{1}{%
\alpha }}\right) \right) }\cdot \frac{\tilde{c}_{\varepsilon }^{k}}{k!}e^{-%
\tilde{c}_{\varepsilon }}d\varepsilon  \label{1}
\end{equation}
with 
\begin{equation}
\tilde{c}_{\varepsilon }=\frac{k_{0}\cdot A}{\left[ \varphi \left( \varphi
^{-1}\left( \left( \frac{k_{0}\cdot A}{k+\varepsilon }\right) ^{\frac{1}{%
\alpha }}\right) \right) \right] ^{\alpha }}=k+\varepsilon\ .
\end{equation}
Here we used the relation $D\varphi ^{-1}\left( x\right) =\left[ \left(
D\varphi \right) \left( \varphi ^{-1}\left( x\right) \right) \right] ^{-1}$
. Using Stirling's formula $k!=\left( 2\pi k\right) ^{\frac{1}{2}}\left( 
\frac{k}{e}\right) ^{k}\left( 1+o\left( 1\right) \right) $ and the
approximation $\left( 1+\frac{\varepsilon }{k}\right) ^{k}=e^{\varepsilon -%
\frac{\varepsilon ^{2}}{2k}+o\left( 1\right) }$ for $\left| \varepsilon
\right| \ll k^{\frac{2}{3}}$ we get for the Poisson term in the integral 
\begin{equation}
\frac{\tilde{c}_{\varepsilon }^{k}}{k!}e^{-\tilde{c}_{\varepsilon }}=\left(
2\pi k\right) ^{-\frac{1}{2}}\cdot e^{-\frac{\varepsilon ^{2}}{2k}+o\left(
1\right) }\ .
\end{equation}
Since $\left| \varepsilon \right| \ll k$ we obtain from formula (\ref{1}) 
\begin{eqnarray}
\lefteqn{\psi _{ess}\left( k\right) =}\nonumber \\
&&\!\!\!\!\!\frac{\left( k_{0}\cdot A\right) ^{\frac{1}{%
\alpha }}\left( 1+o\left( 1\right) \right) }{\alpha k^{1+\frac{1}{\alpha }}}%
\cdot \int\limits_{-k^{\frac{1}{2}+\gamma }}^{k^{\frac{1}{2}+\gamma }}\frac{%
-\varphi \circ \varphi ^{-1}\left( \left( \frac{k_{0}\cdot A}{k+\varepsilon }%
\right) ^{\frac{1}{\alpha }}\right) }{\left( D\varphi \right) \left( \varphi
^{-1}\left( \left( \frac{k_{0}\cdot A}{k+\varepsilon }\right) ^{\frac{1}{%
\alpha }}\right) \right) }\cdot \frac{e^{-\frac{\varepsilon ^{2}}{2k}%
+o\left( 1\right) }}{\left( 2\pi k\right) ^{\frac{1}{2}}}d\varepsilon
\label{2}
\end{eqnarray}
If the decay of $\varphi $ is faster then any power-law we will show that 
\begin{equation}
\frac{-\varphi \circ \varphi ^{-1}\left( \left( \frac{k_{0}\cdot A}{%
k+\varepsilon }\right) ^{\frac{1}{\alpha }}\right) }{\left( D\varphi \right)
\left( \varphi ^{-1}\left( \left( \frac{k_{0}\cdot A}{k+\varepsilon }\right)
^{\frac{1}{\alpha }}\right) \right) }=\left( k+\varepsilon \right) ^{o\left(
1\right) }  \label{22}
\end{equation}
from which we can conclude that $\psi _{ess}\left( k\right) =\frac{1}{k^{1+%
\frac{1}{\alpha }+o\left( 1\right) }}$ holds. But first we will deal with
the simpler case when $\varphi $ is itself of power-law type. Let us
assume that $\varphi $ has an exact power-law tail of the form $\frac{C}{%
\omega ^{\beta }}$ . Therefore $\varphi ^{-1}\left( \omega \right) =\left( 
\frac{C}{\omega }\right) ^{\frac{1}{\beta }}$ and 
\begin{equation}
\left( D\varphi \right) \varphi ^{-1}\left( \left( \frac{k_{0}\cdot A}{%
k+\varepsilon }\right) ^{\frac{1}{\alpha }}\right) =\frac{-\beta C^{\frac{-1%
}{\beta }}\left( k_{0}\cdot A\right) ^{\frac{\beta +1}{\alpha \beta }}}{%
\left( k+\varepsilon \right) ^{\frac{\beta +1}{\alpha \beta }}}
\end{equation}
and finally 
\begin{equation}
\frac{-\varphi \circ \varphi ^{-1}\left( \left( \frac{k_{0}\cdot A}{%
k+\varepsilon }\right) ^{\frac{1}{\alpha }}\right) }{\left( D\varphi \right)
\left( \varphi ^{-1}\left( \left( \frac{k_{0}\cdot A}{k+\varepsilon }\right)
^{\frac{1}{\alpha }}\right) \right) }=\frac{C^{\frac{1}{\beta }}\left(
Ak_{0}\right) ^{\frac{-1}{\alpha \beta }}}{\beta \left( k+\varepsilon
\right) ^{\frac{-1}{\alpha \beta }}}\ .
\end{equation}
Since 
\begin{equation}
\int\limits_{-k^{\frac{1}{2}+\gamma }}^{k^{\frac{1}{2}+\gamma }}\frac{e^{-%
\frac{\varepsilon ^{2}}{2k}+o\left( 1\right) }}{\left( 2\pi k\right) ^{\frac{%
1}{2}}}d\varepsilon \simeq 1+o\left( 1\right)
\end{equation}
we get for the essential part of the degree distribution 
\begin{equation}
\psi _{ess}\left( k\right) =\frac{C^{\frac{1}{\beta }}\left( k_{0}\cdot
A\right) ^{\frac{1}{\alpha }-\frac{1}{\alpha \beta }}\left( 1+o\left(
1\right) \right) }{\beta \alpha k^{1+\frac{1}{\alpha }-\frac{1}{\alpha \beta 
}}}=\frac{1}{k^{1+\frac{1}{\alpha }-\frac{1}{\alpha \beta }+o\left( 1\right)
}}\ .
\end{equation}
The correction term coming from the integration outside the essential
domain can be estimated as follows 
\begin{align}
\psi _{small}\left( k\right) & <const\cdot \left( \int\limits_{-k}^{-k^{%
\frac{1}{2}+\gamma }}\frac{e^{-\frac{\varepsilon ^{2}}{2k}+o\left( 1\right) }%
}{\left( 2\pi k\right) ^{\frac{1}{2}}}d\varepsilon +\int\limits_{k^{\frac{1}{%
2}+\gamma }}^{\infty }\frac{e^{-\frac{\varepsilon ^{2}}{2k}+o\left( 1\right)
}}{\left( 2\pi k\right) ^{\frac{1}{2}}}d\varepsilon \right) \nonumber \\
& =o\left( \frac{1}{k^{1+\frac{1}{\alpha }-\frac{1}{\alpha \beta }}}\right)\ .
\end{align}
In the case when $\varphi \left( \omega \right) $ is a power law in our
weaker sense, say $\varphi \left( \omega \right) =\frac{1}{\omega ^{\beta
+o\left( 1\right) }}$, we need some regularity condition of the form $%
D\varphi \left( \omega \right) =\frac{-1}{\omega ^{1+\beta +o\left( 1\right)
}}$ on the derivative to get an asymptotic power law for $\psi \left(
k\right)$.

We continue now with the case when $\varphi \left( \omega \right) $ decays
faster then any power law, that is we assume 
\begin{equation}
\varphi \left( \omega \right) <\frac{1}{\omega ^{l}}\text{ for any }l\text{
and }\omega >\omega _{0}\left( l\right)\ .
\end{equation}
To get the desired result we have to show that formula (\ref{22}) holds.
Since $\varphi ^{-1}\left( \frac{1}{\omega }\right) $ goes to infinity for $%
\omega \rightarrow \infty $ we have to show 
\begin{equation}
\frac{-\varphi \left( x\right) }{D\varphi \left( x\right) }=\left[ \varphi
\left( x\right) \right] ^{o_{x}\left( 1\right) }\ .  \label{111}
\end{equation}
The last formula states that the negative logarithmic derivative of $%
\varphi $ should not become to large or to small compared to $\varphi $
respectively $\frac{1}{\varphi }$. For the following it is convenient to
set $\varphi \left( x\right) =e^{-g\left( x\right) }$ with $g\left( x\right)
\rightarrow \infty $ and rewrite formula (\ref{111}) as 
\begin{equation}
e^{-\mu g\left( x\right) }<\frac{1}{Dg\left( x\right) }<e^{\mu g\left(
x\right) }\text{ for }\mu \in \left( 0,\mu _{0}\right) \text{ and }%
x>x_{0}\left( \mu \right)\ .  \label{43}
\end{equation}
Assume that formula (\ref{43}) is not true with respect to the right hand
side. Than we have for a sequence of values $\left\{ x_{i}\right\} $ and
open intervals $I_{i}$ around the $x_{i}$ and some function $a\left(
x\right) $ 
\begin{equation}
\frac{1}{Dg\left( x\right) }=e^{\mu g\left( x\right) }a\left( x\right) \text{
and }a\left( x\right) >1\text{ for }x\in I_{i} \ . \label{33}
\end{equation}
Integrating the last equation gives 
\begin{equation}
e^{\mu g\left( x\right) }=e^{\mu g\left( x_{0}\right) }+\mu
\int\limits_{x_{0}}^{x}\frac{1}{a\left( z\right) }dz\ .
\end{equation}
Since our assumption on $D\left[ \frac{1}{\varphi \left( \omega \right) }%
\right] ^{\mu }$ to be monotone for $\mu >0$ and $\omega >\omega _{0}\left(
\mu \right) $ implies $a\left( x\right) >1$ eventually we conclude that 
\begin{equation}
e^{\mu g\left( x\right) }<e^{\mu g\left( x_{0}\right) }+\mu \left(
x-x_{0}\right) \ . \label{44}
\end{equation}
But the fast decay condition for $\varphi \left( x\right) $ translates
into a growth condition for $g\left( x\right) $ namely for all $k$ 
\begin{equation}
g\left( x\right) >k\log x;\text{ }x>x_{0}\left( k\right)
\end{equation}
which clearly contradicts formula (\ref{44}). It remains to show that the
left hand side of formula (\ref{43}) also holds. Assuming the converse we
get 
\begin{equation}
\frac{1}{Dg\left( x\right) }=e^{-\mu g\left( x\right) }\frac{1}{a\left(
x\right) }\text{ and }a\left( x\right) >1\text{ for }x\in I_{i}
\end{equation}
and after integration 
\begin{equation}
e^{-\mu g\left( x\right) }=e^{-\mu g\left( x_{0}\right) }-\mu
\int\limits_{x_{0}}^{x}a\left( z\right) dz\ .
\end{equation}
The monotonicity condition again implies $a\left( x\right) >1$ eventually,
hence 
\begin{equation}
e^{-\mu g\left( x\right) }<e^{-\mu g\left( x_{0}\right) }-\mu \left(
x-x_{0}\right)
\end{equation}
and a clear contradiction since the right hand side becomes negative for
large values of $x$. With this result in hand we get for the leading term 
\begin{eqnarray}
\psi _{ess}\left( k\right) & =\frac{\left( k_{0}A\right) ^{\frac{1}{\alpha }%
}\left( 1+o\left( 1\right) \right) }{\alpha k^{1+\frac{1}{\alpha }}}\cdot
\int\limits_{-k^{\frac{1}{2}+\gamma }}^{k^{\frac{1}{2}+\gamma }}\left( \frac{%
k_{0}A}{k+\varepsilon }\right) ^{\frac{o\left( 1\right) }{\alpha }}\cdot 
\frac{e^{-\frac{\varepsilon ^{2}}{2k}+o\left( 1\right) }}{\left( 2\pi
k\right) ^{\frac{1}{2}}}d\varepsilon\nonumber \\
& =\frac{1}{k^{1+\frac{1}{\alpha }+o\left( 1\right) }}\int\limits_{-k^{\frac{%
1}{2}+\gamma }}^{k^{\frac{1}{2}+\gamma }}\frac{e^{-\frac{\varepsilon ^{2}}{2k%
}+o\left( 1\right) }}{\left( 2\pi k\right) ^{\frac{1}{2}}}d\varepsilon =%
\frac{1}{k^{1+\frac{1}{\alpha }+o\left( 1\right) }}\ .
\end{eqnarray}
For the minor integrals we get 
\begin{equation}
\psi _{small}^{\left( 1\right) }\left( k\right) =\int\limits_{-k}^{-k^{\frac{%
1}{2}+\gamma }}\frac{\left( k_{0}A\right) ^{\frac{1}{\alpha }}}{\alpha
\left( k+\varepsilon \right) ^{1+\frac{1}{\alpha }}}\cdot \frac{-\varphi
\circ \varphi ^{-1}\left( \left( \frac{k_{0}A}{k+\varepsilon }\right) ^{%
\frac{1}{\alpha }}\right) }{\left( D\varphi \right) \left( \varphi
^{-1}\left( \left( \frac{k_{0}\cdot A}{k+\varepsilon }\right) ^{\frac{1}{%
\alpha }}\right) \right) }\cdot \frac{e^{-\frac{\varepsilon ^{2}}{2k}%
+o\left( 1\right) }}{\left( 2\pi k\right) ^{\frac{1}{2}}}d\varepsilon
\end{equation}
and 
\begin{eqnarray}
\psi _{small}^{\left( 2\right) }\left( k\right) & =\int\limits_{k^{\frac{1}{2%
}+\gamma }}^{\infty }\frac{\left( k_{0}A\right) ^{\frac{1}{\alpha }}}{\alpha
\left( k+\varepsilon \right) ^{1+\frac{1}{\alpha }}}\cdot \left( \frac{k_{0}A%
}{k+\varepsilon }\right) ^{\frac{o\left( 1\right) }{\alpha }}\cdot \frac{e^{-%
\frac{\varepsilon ^{2}}{2k}+o\left( 1\right) }}{\left( 2\pi k\right) ^{\frac{%
1}{2}}}d\varepsilon\nonumber \\
& =o\left( \frac{1}{k^{1+\frac{1}{\alpha }}}\right)\ .
\end{eqnarray}
To see that $\psi _{small}^{\left( 1\right) }\left( k\right) $ is of order 
$o\left( \frac{1}{k^{1+\frac{1}{\alpha }}}\right) $ fix $\gamma >\delta >0$
sufficiently small and some $k^{\prime }$ such that 
\begin{equation}
\frac{-\varphi \circ \varphi ^{-1}\left( \left( \frac{k_{0}A}{k+\varepsilon }%
\right) ^{\frac{1}{\alpha }}\right) }{\left( D\varphi \right) \left( \varphi
^{-1}\left( \left( \frac{k_{0}\cdot A}{k+\varepsilon }\right) ^{\frac{1}{%
\alpha }}\right) \right) }<\left( \frac{k}{k_{0}A}\right) ^{\frac{\delta }{%
\alpha }}\text{ for }k>k^{\prime }\ .
\end{equation}
With 
\begin{equation}
C_{\max }\left( \delta _{0}\right) =\max\limits_{0<k<k^{\prime }}\left\{ 
\frac{-\varphi \circ \varphi ^{-1}\left( \left( \frac{k_{0}A}{k}\right) ^{%
\frac{1}{\alpha }}\right) }{\left( D\varphi \right) \left( \varphi
^{-1}\left( \left( \frac{k_{0}A}{k}\right) ^{\frac{1}{\alpha }}\right)
\right) }\right\} <\infty
\end{equation}
we obtain for $\psi _{small}^{\left( 1\right) }\left( k\right) $ the
following estimation: 
\begin{eqnarray}
\psi _{small}^{\left( 1\right) }\left( k\right) & <\left( k_{0}A\right) ^{%
\frac{1}{\alpha }}\cdot C_{\max }\left( \delta \right) k^{\prime }e^{-\left( 
\frac{k}{2}-2k^{\prime }\right) }+\nonumber \\
& +\int\limits_{k^{\prime }}^{k-k^{\frac{1}{2}+\gamma }}\frac{\left(
k_{0}A\right) ^{\frac{1}{\alpha }}}{\alpha \varepsilon ^{1+\frac{1}{\alpha }}%
}\cdot \left( \frac{\varepsilon }{k_{0}A}\right) ^{\frac{\delta }{\alpha }%
}\cdot \frac{e^{-\frac{k^{2\gamma }}{2}+o\left( 1\right) }}{\left( 2\pi
k\right) ^{\frac{1}{2}}}d\varepsilon \nonumber \\
& =o\left( \frac{1}{k^{1+\frac{1}{\alpha }}}\right)\ .
\end{eqnarray}
Therefore we obtain in the case when $\varphi $ decays faster then any
power law that the degree distribution $\psi \left( k\right) =\frac{1}{k^{1+%
\frac{1}{\alpha }+o\left( 1\right) }}$ is independent of $\varphi .$ $%
\square $

\subsection{The case of nonconstant outdegree}

Next we deal with the case when the outdegree distribution is not constant.
We will always assume that the expected outdegree is finite to keep the
total number of edges proportional to $N.$ First we show that the indegree
distribution is the same as in the constant degree case with $k_{0}$
replaced by ${\Bbb E}\left( d_{out}\left( x\right) \right)$. With respect
to the total degree distribution two situations can appear. Either the r.v. $%
d_{out}$ is independent of the r.v. $\omega$ or not. In the independent
case one has to take then just the convolution of indegree and outdegree
distribution to get the total degree distribution. The guiding principle
here is, that the dominating distribution wins. If things are not
independent the analysis becomes quite more involved and we will sketch only
the relevant steps dealing with a detailed investigation in a separate paper.

We start with the independent case. Let 
\begin{equation}
q_{k}=\Pr\left\{ d_{out}\left( x\right) =k\right\}
\end{equation}
and 
\begin{equation}
p_{k}=\Pr\left\{ d_{in}\left( x\right) =k\right\}
\end{equation}
the asymptotic outdegree respectively indegree distributions. Since we
assumed asymptotic independence we get for the total degree distribution $%
d_{k}$ the convolution 
\begin{equation}
d_{k}=\sum\limits_{0\leq i\leq k}q_{i}\cdot p_{k-i}\ .
\end{equation}
We say that the tail of a distribution $\left\{ p_{i}\right\} $ dominates
the tail of another distribution $\left\{ q_{i}\right\} $ if $q_{i}=o\left(
p_{i}\right).$ For the $\left\{ p_{i}\right\} $ we assume further that it
is an asymptotic power-law distribution with exponent $\gamma>2.$ We
estimate $d_{k}$ as follows: 
\begin{eqnarray}
d_{k} & =\sum\limits_{0\leq i\leq\frac{k}{2}}q_{i}\cdot
p_{k-i}+\sum\limits_{0\leq i<\frac{k}{2}}q_{k-i}\cdot p_{i}\nonumber \\
& =\sum\limits_{0\leq i\leq\frac{k}{2}}q_{i}\cdot\frac{1}{\left( k-i\right)
^{\gamma+o\left( 1\right) }}+\sum\limits_{0\leq i<\frac{k}{2}}o\left( \frac{1%
}{\left( k-i\right) ^{\gamma+o\left( 1\right) }}\right) \cdot \frac{1}{%
i^{\gamma+o\left( 1\right) }} \nonumber \\
& =\frac{1}{k^{^{\gamma+o\left( 1\right) }}}\sum\limits_{0\leq i\leq \frac{k%
}{2}}q_{i}\cdot\frac{1}{\left( 1-\frac{i}{k}\right) ^{\gamma+o\left(
1\right) }}+o\left( \frac{1}{k^{^{\gamma+o\left( 1\right) }}}\right)
\sum\limits_{0\leq i<\frac{k}{2}}\frac{1}{\left( 1-\frac{i}{k}\right)
^{\gamma+o\left( 1\right) }}\cdot p_{i}\ .
\end{eqnarray}
Since both sums in the last expression are bounded and larger zero we
obtain 
\begin{equation}
d_{k}=\frac{1}{k^{\gamma+o\left( 1\right) }}\ .
\end{equation}
In the case when the outdegree distribution is the dominating one ( and in
our case it has to be a power-law distribution) we can just reverse the
above argument to show that the total degree distribution is in this
situation the outdegree distribution.

It remains to show that a variable outdegree will not affect the structure
of the indegree distribution. The main difference in comparison to the
previous section is that formula (\ref{5}) gets replaced by 
\begin{eqnarray}
\lefteqn{\psi _{\omega }\left( k\right)= }  \nonumber\\
& & \sum\limits_{M}\Pr \left\{ \sharp \text{ of
edges in }V_{N}=M\right\} \binom{M}{k}\left( \frac{A\left( \xi ,N\right) }{%
\left[ \varphi \left( \omega \right) \right] ^{\alpha }N}\right) ^{k}\left(
1-\frac{A\left( \xi ,N\right) }{\left[ \varphi \left( \omega \right) \right]
^{\alpha }N}\right) ^{M-k}\ .
\end{eqnarray}
Since for $M=const\cdot N$ each binomial term still converges to a Poisson
distribution with parameter $\frac{const\cdot A}{\left[ \varphi \left(
\omega \right) \right] ^{\alpha }}$ it is not difficult to show that in case
when a limiting outdegree distribution exists and the first moment is finite
one still obtains a limiting Poisson distribution where $k_{0}$ is replaced
by ${\Bbb E}\left( d_{out}\right).$ 

\subsection{The degree correlation and the clustering coefficient}

It is interesting to note that graphs formed by the ''Cameo Principle''\
show a degree affinity which is observed in many real networks and which is
also one of the underlying assumption in the Albert Barabasi model, namely 
\begin{equation}
\Pr \left\{ \text{an edge starting in }x\text{ ends up in a vertex }y\text{
with }d\left( y\right) =k\right\} =\frac{k}{N}\cdot const\ .
\end{equation}
To see that this is the case recall from the previous sections that the
typical vertices with degree $k$ have $\omega -$ values from $I_{ess}\left(
k\right) $ for large $k$. Due to the definition of the directed edge
probability we get for $\left| \varepsilon \right| <k^{\frac{1}{2}+\gamma }$ 
\begin{eqnarray}
\Pr \left\{ x\rightarrow y\mid \omega \left( y\right) \in I_{ess}\left(
k\right) \right\} & =\frac{A}{\left[ \varphi \left( \varphi ^{-1}\left(
\left( \frac{k_{0}A}{k+\varepsilon }\right) ^{\frac{1}{\alpha }}\right)
\right) \right] ^{\alpha }}\cdot \frac{1}{N}\text{ } \nonumber \\
& =\frac{A}{\left( \frac{k_{0}A}{k+\varepsilon }\right) }\cdot \frac{1}{N}=%
\frac{k\left( 1+o\left( 1\right) \right) }{k_{0}N}=\frac{k\left( 1+o\left(
1\right) \right) }{\sum\limits_{x\in V_{N}}d\left( x\right) } \nonumber \\
& =\Pr \left\{ x\rightarrow y\mid d\left( y\right) =k\right\}\ .
\end{eqnarray}
Therefore it seems to us, that if one observes in real networks edge
affinities proportional to the degree one cannot conclude that the
generation of the graph is also due to an edge formation process build up
from this proportionality (like in the Albert-Barabasi model).

We want to close this section with a short discussion of the clustering
coefficient. Since no explicit local clustering rule was implemented in our
model one cannot expect a high clustering coefficient. Nevertheless it will
be slightly larger then in the classical $G\left( N,p\right)$ case. Again
we will study here only the situation for constant outdegree $k_{0}$.
Remember that the clustering coefficient of a vertex $x$ with degree $k$ is
defined as 
\begin{equation}
C\left( x\right) =\frac{2\cdot \sharp \left\{ \left( y,z\right) ;y\sim z%
\text{ and }y,z\in N_{1}\left( x\right) \right\} }{k\left( k-1\right) }
\end{equation}
where $N_{1}\left( x\right) $ is the set of neighbors of $x$. The global
clustering coefficient is then just the expectation ${\Bbb E}\left( C\left(
x\right) \right).$ In what follows we will only give an explicit
computation for the number of edges in $N_{1}\left( x\right) $ which is a
little bit simpler to compute. Since the edge formation probabilities depend
only on the ''target''\ vertices, we get on $N_{1}\left( x\right) $ the same 
$\omega -$ distribution as on the whole graph. In other words for any fixed $%
x$ the probability that a $y\in N_{1}\left( x\right) $ has $\omega \left(
y\right) >\omega _{0}$ is given by $\int\limits_{\omega _{0}}^{\infty
}\varphi \left( \omega \right) d\omega $. Therefore to compute ${\Bbb E}%
\left( \sharp \left\{ \left( y,z\right) ;y\sim z\text{ and }y,z\in
N_{1}\left( x\right) \right\} \mid d\left( x\right)=\right.$ $\left. =k\right) $ it is enough
to compute the expected number of edges in a random vertex sample of given
size $k.$ It is easy to see that this number equals $\frac{k_{0}\left(
k-1\right) }{N}.$ Since we know further that in the limit $N\rightarrow
\infty $ the degree distribution for an $x$ with $\omega \left( x\right)
=\omega $ is the Poisson distribution $\frac{c_{\omega }^{k-k_{0}}}{\left(
k-k_{0}\right) !}e^{-c_{\omega }}$ with $c_{\omega }=\frac{k_{0}\cdot A}{%
\left[ \varphi \left( \omega \right) \right] ^{\alpha }}$ and $k\geq k_{0}$
we get with $T\left( \omega \right) :={\Bbb E}\left( \sharp \left\{ \left(
y,z\right) ;y\sim z\text{ and }y,z\in N_{1}\left( x\right) \right\} \mid
\omega \left( x\right) =\omega \right) $ 
\begin{eqnarray}
T\left( \omega \right) & =\sum\limits_{k\geq k_{0}}\frac{k_{0}\left(
k-1\right) c_{\omega }^{k-k_{0}}}{N\left( k-k_{0}\right) !}e^{-c_{\omega }}
\nonumber \\
& =\sum\limits_{m\geq 0}\frac{k_{0}\left( m+k_{0}-1\right) c_{\omega }^{m}}{%
N\cdot m!}e^{-c_{\omega }}\nonumber \\
& =\frac{k_{0}c_{\omega }+k_{0}\left( k_{0}-1\right) }{N}
\end{eqnarray}
and finally 
\begin{equation}
\int\limits_{\omega }T\left( \omega \right) d\omega =\int \frac{k_{0}\frac{%
k_{0}\cdot A}{\left[ \varphi \left( \omega \right) \right] ^{\alpha }}%
+k_{0}\left( k_{0}-1\right) }{N}d\omega
\end{equation}
as the expected number of edges among the neighbors of a random chosen
vertex $x$. Completely analogous one can obtains the asymptotics $\frac{const%
}{N}$ for the expected clustering coefficient $C.$ The explicit expression
for the constant there is more cumbersome but straightforward.

\section{A growing network model}

In the above sections we described a model where the total number of
vertices is large but fixed and all edges are generated at once
simultaneously. Since real networks are typically evolving networks many
models allow a growth process for the graphs. This is easy to incorporate in
our situation too. For simplicity and since it is rather common today we use
the evolution concept from Albert and Barabasi \cite{3}.

We start with an initial graph $G_{0}$ with $N_{0}$ vertices and at each
time $t\in {\Bbb N}$ a fixed number $L_{0}$ of new vertices with $\omega $-
values taken again i.i.d. with density $\varphi \left( \omega \right) .$
Each vertex forms at the time when it enters the graph, $k_{0}$ edges with
the vertices already present. The edges are formed according to the ''Cameo
Principle'' that is 
\begin{eqnarray}
p_{\omega }\left( t\right) &:=&
\Pr \left\{ x\rightarrow y\mid \omega = 
\omega
\left( y\right) \text{ and }x\text{ enters the graph at time }t\right\} =\nonumber \\
&=& \frac{A\left( \xi _{t},N_{t}\right) }{\left[ \varphi \left( \omega \right) %
\right] ^{\alpha }}\cdot \frac{1}{N_{t}}
\end{eqnarray}
where $\xi _{t}$ is the realization of the $\omega $- values at time $t$, $%
N_{t}=tL_{0}+N_{0}$ and $A\left( \xi _{t},N_{t}\right) $ a normalization
constant. Let $d\left( x,t,\tau \right) $ be the degree of $x$ at time $t$
and $\tau \left( x\right) =\tau $ the time when $x$ entered the graph. For $%
x $ fixed and $t\gg \tau $ we have 
\begin{equation}
{\Bbb E}\left( d_{in}\left( x,t\right) \right) =\sum\limits_{i>\tau }^{t}%
\frac{k_{0}A\left( \xi _{i},N_{i}\right) }{\left[ \varphi \left( \omega
\left( x\right) \right) \right] ^{\alpha }}\cdot \frac{1}{N_{t}}\simeq \frac{%
k_{0}A}{\left[ \varphi \left( \omega \left( x\right) \right) \right]
^{\alpha }\cdot L_{0}}\left( \log t-\log \tau \right)\ .
\end{equation}
We now investigate the asymptotic indegree distribution, namely the
asymptotic probability that a randomly chosen vertex from the evolving graph
at time $t$ has indegree $k.$ For simplicity we deal here only with the case 
$L_{0}=1$. We first look at the indegree distribution $\psi _{\omega }\left(
k,\tau ,t\right) =$ $\Pr \left\{ d_{in}\left( x,t\right) =k\right\} $ for a
fixed $x$ with $\omega =\omega \left( x\right) $ and $\tau =\tau \left(
x\right)$. Since 
\begin{equation}
\psi _{\omega }\left( k,\tau ,t\right) =\Pr \left\{ \sum\limits_{y:\tau
\left( y\right) >\tau \left( x\right) }{\bf 1}_{y\sim x}=k\right\}
\end{equation}
with ${\bf 1}_{y\sim x}$ being the independent r.v. of the indicator
function that $y$ formed an edge with $x$ , one gets by a well known theorem
in probability theory for $t-\tau $ large a limiting Poisson distribution
with expectation equal to ${\Bbb E}\left( d_{in}\left( x,t\right) \right) $
(see \cite{44} for recent sharp results in this direction)$.$ In complete
analogy to section 2 we obtain after integration over $\omega $ a scale free
distribution for $\psi \left( k,\tau ,t\right) $- the not on $\omega $
conditioned indegree distribution. Taking explicit care on the $\tau $
dependence we get 
\begin{equation}
\psi \left( k,\tau ,t\right) =\frac{\left( k_{0}A\left[ \log t-\log \tau %
\right] \right) ^{\frac{1}{\alpha }}}{k^{1+\frac{1}{\alpha }+o\left(
1\right) }};t-\tau >\frac{k}{k_{0}}\ .
\end{equation}
Note that $o\left( 1\right) $ does not depend on $\tau $ for $t-\tau \gg
k. $ To obtain the final (not on $\tau $ conditioned) distribution $\psi
\left( k,t\right) $ we just have to sum 
\begin{eqnarray}
\psi \left( k,t\right) &=&\frac{1}{t}\sum\limits_{0<\tau <t}\psi \left(
k,\tau ,t\right) \nonumber \\
&=&\frac{1}{t}\sum\limits_{0<\tau <t}\frac{\left( k_{0}A\left[ \log t-\log
\tau \right] \right) ^{\frac{1}{\alpha }}}{k^{1+\frac{1}{\alpha }+o\left(
1\right) }} \nonumber \\
&=&\frac{1}{k^{1+\frac{1}{\alpha }+o\left( 1\right) }}\ .
\end{eqnarray}
In the last two equalities we have implicitly used the fact that the terms
with $t-\tau \sim const\cdot k$ do not really contribute to the $\psi \left(
k,t\right) $ distribution and that $\sum\limits_{0<\tau <t}\log \tau \sim
t\log t-t$.

\section{A model variant with high clustering coeffi\-cient (the ''my friends
are your friends''--prin\-ciple)}

To get high clustering coefficients one has to incorporate local edge
formation rules into the model like in the famous small world model of Watts
and Strogatz \cite{4}. A natural way to do this for our graphs is by
introducing a kind of additional local choice rule: with probability $p$
make the choice into the neighbors of your own neighbors at the given time
and with probability $q=1-p$ proceed like above. We call this rule the ''my
fiends are your friends'' principle for obvious reasons. It should be a well
known principle in social sciences since many social contacts are formed
that way. We will give a detailed investigation of the types of graphs
resulting from such processes in a forthcoming paper \cite{55}. Here we will
present only an informal discussion of some of the main properties. First it
is important to note that the order in which the edges are created plays now
a role (at least for the fine structure of local properties of the graphs).
To fix the setting let us consider the following rule. The enumeration of
the vertex set we will associate with the time at which the vertex will make
a choice. The first choice each vertex makes is an independent one according
to the rules specified in the previous paragraph. For the next choices any
vertex $x$ has two options -- with probability $q=1-p$ he proceeds as with
the first choice (independent, global choice) or with probability $p$ he
chooses from $N_{2}\left( x,t\right)$, the set of vertices at distance $2$
from $x$ at time $t$ (dependent, local choice). We remark that the concrete
form of $\varphi \left( \omega \right) $ is irrelevant for the further
discussion. Two situations are now natural to consider:

Case A: At an initial set of choices all vertices make their first,
independent choice (here the order in which the choices are made does not
matter). After that, each vertex makes it's second choice at the time
corresponding to it's index number, then all vertices make their third
choice and so on (to avoid a bias one should renumber randomly the vertex
set after each sequence of choices).

Case B: Each vertex makes the remaining $k_{0}-1$ choices immediately after
the first one. Again the time when a vertex is about to make the choices
corresponds to it's index number.

Let $T\left( x\right) $ be the number of triangles in the final graph
containing $x$. There are some trivial lower and upper bounds on the
expectation of $T\left( x\right).$ First observe that due to the bounded
outdegree there cannot be more than $\left( k_{0}-1\right) d\left( x\right) $
edges among $N_{1}\left( x\right)$. On the other side any choice made by $%
x $ according to the ''my friends are your friends''-rule automatically
generates a triangle containing $x.$ Therefore ${\Bbb E}\left( T\left(
x\right) \right) \geq p\left( k_{0}-1\right) $ and hence the correlation
coefficient is independent of $N$ asymptotically. We claim that for large $%
d\left( x\right) $ the conditional expectation given the degree is
approximately ${\Bbb E}\left( T\left( x\right) \mid d\left( x\right) \right)
\simeq const\cdot d\left( x\right).$ We outline the reason for that for $%
k_{0}=2$ and the situation in case B. Note that the rule from case B is also
the canonical one if one wants to deal with growing networks.

Let $d_{in}^{1}\left( x\right) $ be the indegree of $x$ after each vertex
made the first choice. Denote further by $N_{l}^{1}\left( x\right) $ the set
of vertices at distance $l$ from $x$ after the first choices are made. Since
the first choices are independent, a vertex $x$ is not part of a triangle
almost surely. In the second round of choices triangular configurations will
be formed frequently. Since we assumed that $d\left( x\right) $ is large we
can assume that $d_{in}^{1}\left( x\right) $ is large too. There are two
mechanisms which contribute to $T\left( x\right) $ when the second choices
are made. First all vertices in $N_{1}^{1}\left( x\right) $ create an edge
into $N_{l}^{1}\left( x\right) $ with probability $p$ and since $\left|
N_{1}^{1}\left( x\right) \right| =d_{in}^{1}\left( x\right) +1$ is assumed
to be large we can neglect the events that multiple edges are created. So we
expect $p(d_{in}^{1}\left( x\right) +1)$ triangles containing $x$ from this
process. Second any vertex $z$ from $N_{2}^{1}\left( x\right) $ can chose $x$
with probability $\frac{p}{\left| N_{2}^{1}\left( z\right) \right| }$ and
therefore create a new triangle. Furthermore for most vertices $z\in
N_{2}^{1}\left( x\right) $ we have 
\begin{equation}
\left| N_{2}^{1}\left( x\right) \right| \sim d_{in}^{1}\left( x\right) \cdot
const\cdot \left| N_{2}^{1}\left( z\right) \right|
\end{equation}
from which we get an estimation of order $pd_{in}^{1}\left( x\right) \cdot
const$ for the expected number of triangles generated via this process.
Combining this estimations and using the fact that the number of independent
choices pointing to $x$ in the second choice sequence is about $\left(
1-p\right) d_{in}^{1}\left( x\right) $ we obtain 
\begin{eqnarray}
d\left( x\right) &\sim &d_{in}^{1}\left( x\right) +\left( 1-p\right)
d_{in}^{1}\left( x\right) +p(d_{in}^{1}\left( x\right) +1)+pd_{in}^{1}\left(
x\right) \cdot const \nonumber \\
&\sim &d_{in}^{1}\left( x\right) \left( 2+p\cdot const\right)
\end{eqnarray}
which gives for the expected number of triangles 
\begin{equation}
{\Bbb E}\left( T\left( x\right) \mid d\left( x\right) \right) \sim \frac{%
p\left( 1+const\right) }{2+p\cdot const}d\left( x\right)\ .
\end{equation}
The above considerations are of course of heuristic type but can easily
made be precise. The $const$ appearing in the above formulas can as well be
explicitly computed. We remark finally that the above considerations remain
true if the outdegree distribution is not a constant and that the clustering
mechanism does not change the asymptotic shape of the total degree
distribution. We plan to discuss all this aspects in much more detail in a
paper in preparation.

\section{Conclusions and outlook}

We have shown in this article that graphs with a scale-free degree
distribution appear naturally as the result of a simple edge formation rule
based on choices where the probability to chose a vertex with affinity
parameter $\omega $ is proportional to the frequency of appearance of that
parameter. If the affinity parameter $\omega $ is itself power law like
distributed one could also use a direct proportionality to the value $\omega 
$ to get still a scale free graph since $\varphi \left( \omega \right) $ is
itself a polynomial relation (this explains the observation reported in \cite
{2}).

We have not tried in this paper to give estimations on the error terms for
the asymptotics. Numerical simulations, which will be reported elsewhere,
indicate that already for values of $k$ of order $const\cdot k_{0}$ and
vertex sets of size $N\approx 10^{5}$ one has a very good agreement with the
asymptotic result. Of course in all this finite size effects the concrete
form of $\varphi $ really matters.

It seems natural to investigate models where $\varphi \left( \omega \right) $
has a singular, for instance discrete, support. Such a situation is common
when one couples for instance the choice probabilities directly to the
outdegree (see \cite{33} for a specific model of such type). Some care is
necessary in that case. For instance if one takes a discrete integer valued
random variable $k$ with the distribution $\varphi \left( k\right)
=const\cdot \exp \left( -k\right) $ and proceeds like in the case of the
continuous variable $\omega $ it will be not longer true that the asymptotic
degree distribution is a power-law distribution. The reason behind this is
the following. The expected indegree of vertices $x$ and $y$ with $k\left(
x\right) =k_{0}$ and $k\left( y\right) =k_{0}+1$ differs by a constant
factor but since the indegree of $x$ and $y$ is concentrated in an interval
of length square root of the expectation around the expectation there
appears a gap in the degree distribution of the whole graph. If $\varphi
\left( k\right) $ is a power law distribution with exponent $\beta $ the
same remains true for $\alpha >\frac{2}{\beta }$ whereas for $\alpha <\frac{2%
}{\beta }$ a similar proof with the same results as in the continuous case
can be given. In general one needs for singular $\varphi \left( \omega
\right) $ an overlap condition for the involved Poisson distributions to get
a power law tail distribution.

With respect to the clustering mechanism described in section 4 there are
many interesting questions. What is the effect on diameter, average
path-length and spectral properties?

{\em Acknowledgment: }We would like to thank Andreas Ruschhaupt and
Madeleine Sirugue-Collin for stimulating and inspiring discussions on
several aspects of the paper.


\end{document}